# A Two-Part Mixed-Effects Modeling Framework For Analyzing Whole-Brain Network Data

Sean L. Simpson

*Department of Biostatistical Sciences, Wake Forest School of Medicine,*

*Winston-Salem, NC 27157-1063*

Paul J. Laurienti

*Department of Radiology, Wake Forest School of Medicine,*

*Winston-Salem, NC 27157-1063*

Sean L. Simpson is Assistant Professor, Department of Biostatistical Sciences, Wake Forest School of Medicine, Winston-Salem, NC 27157 (E-mail: slsimpso@wakehealth.edu). Paul J. Laurienti is Professor, Department of Radiology, Wake Forest School of Medicine, Winston-Salem, NC 27157 (E-mail: plaurien@wfubmc.edu).

# ABSTRACT

Whole-brain network analyses remain the vanguard in neuroimaging research, coming to prominence within the last decade. Network science approaches have facilitated these analyses and allowed examining the brain as an integrated system. However, statistical methods for modeling and comparing groups of networks have lagged behind. Fusing multivariate statistical approaches with network science presents the best path to develop these methods. Toward this end, we propose a two-part mixed-effects modeling framework that allows modeling both the probability of a connection (presence/absence of an edge) and the strength of a connection if it exists. Models within this framework enable quantifying the relationship between an outcome (e.g., disease status) and connectivity patterns in the brain while reducing spurious correlations through inclusion of confounding covariates. They also enable prediction about an outcome based on connectivity structure and vice versa, simulating networks to gain a better understanding of normal ranges of topological variability, and thresholding networks leveraging group information. Thus, they provide a comprehensive approach to studying system level brain properties to further our understanding of normal and abnormal brain function.

# 1. INTRODUCTION

Whole-brain functional magnetic resonance imaging (fMRI) network analyses have moved to the forefront of neuroimaging research over the last decade. fMRI measures localized brain activity by capturing changes in blood flow and oxygenation via the blood oxygen level-dependent (BOLD) contrast (Ogawa et al., 1990). These measurements are recorded from cubic subdivisions of the brain roughly a few millimeters in size called *voxels*. Averaging the BOLD signal time series across voxels within specified regions provides coarser representations. Functional connectivity analysis (FC) examines functional similarities between time series pairs in specified voxels or regions (Sporns, 2010; Biswal et al., 1995; Friston, 1994). Functional brain network analysis serves as a distinct subfield of connectivity analysis in which functional associations are quantified for all $n$ time series pairs to create an interconnected representation of the brain (a brain network). The resulting $n \times n$ connection matrix is generally thresholded to create a binary adjacency matrix that retains "significant" connections while removing weaker ones. Weighted (continuous) network analyses, which we focus on here, have gained traction but still lag behind due to computational and methodological challenges they pose (Telesford et al., 2011; Rubinov and Sporns, 2011; Ginestet et al., 2011). The connection matrix is still often thresholded to remove negative connections (for reasons noted in Telesford et al., 2011; Cao et al., 2014; and others) and/or weak connections in the continuous case. A schematic exhibiting this network generation process is presented in Figure 1.

This emerging area of fMRI brain network analysis allows studying the brain as a system, providing profound clinical insight into the link between system level properties and behavioral and health outcomes (Biswal et al., 2010; Sporns, 2010; Bullmore and Sporns, 2009; Bassett and Bullmore, 2009). The application of network science (an interdisciplinary offshoot of graph theory) has facilitated these analyses and our understanding of how the brain is structurally and functionally organized. Both binary and weighted versions of graph metrics such as degree, clustering coefficient, path length, efficiency, centrality, and modularity serve as common descriptive topological properties of interest. While network science has catalyzed a paradigmatic shift in neuroscience, methods for statistically modeling and comparing groups of networks have lagged behind (Simpson et al., 2013a). These comparisons have great appeal for researchers interested in gaining further insight into complex brain function and how it changes across different mental states and disease conditions. Most current approaches to modeling and comparing brain networks either rely on a specific extracted summary metric (e.g., clustering coefficient) which may lack clinical use due to low sensitivity and specificity, or on mass-univariate nodal or edge-based comparisons that ignore the inherent topological properties of the network while also yielding little power to determine significance (Zalesky et al., 2010; Ginestet et al., 2014). While some univariate approaches like the network-based statistic (NBS) (Zalesky et al., 2010) have proven useful, gleaning deeper insights into normal and abnormal changes in complex functional organization demands methods that leverage the wealth of data present in an entire brain network. This systemic organization confers much of our brains' functional abilities as functional connections may be lost due to an adverse health condition but compensatory connections may develop as a result in order to maintain organizational consistency and functional performance. Consequently, brain network analysis necessitate a multivariate modeling framework that allows assessing the effects of multiple variables of interest and topological network features (e.g., demographics, disease status, nodal clustering, nodal centrality, etc.) on the overall network structure. That is, if we have

$$\text{Data} \begin{cases} \boldsymbol{Y}_i : \text{network of participant } i \\ \boldsymbol{X}_i : \text{covariate information (metrics, demographics, etc.)} \end{cases},$$

we want the ability to model the probability density function of the network given the covariates $P(\boldsymbol{Y}_i | \boldsymbol{X}_i, \boldsymbol{\theta}_i)$, where $\boldsymbol{\theta}_i$ are the parameters that relate the covariates to the network structure.

More recent brain network comparison methods that attempt to better exploit the topological features of network data include the exponential random graph modeling framework (ERGM) (Simpson et al., 2011, 2012), the permutation network framework (PNF) (Simpson et al., 2013b), and the multivariate distance matrix regression (MDMR) framework (Shehzad et al., 2014). While all show promise, they lack the flexibility of the modeling and inferential tools developed for fMRI activation data. The ERGM framework allows efficiently representing complex network data and inherently accounts for higher order dependence/topological properties, but multiple-subject comparisons can pose problems given that these models were originally developed for the modeling of one network at a time (Simpson et al., 2011). Moreover, the amount of programming work increases linearly with the number of subjects since ERGMs must be fitted and assessed for each subject individually (Simpson et al., 2012). Incorporating novel metrics (perhaps more rooted in brain biology) may be difficult due to degeneracy issues that may arise (Handcock, 2002; Rinaldo et al., 2009; O'Malley, 2013). While well-suited for substructural assessments, edge-level examinations remain difficult with these models. Additionally, most ERGM developments have been for binary networks; approaches for weighted networks have been proposed but remain in their infancy (Krivitsky, 2012; Desmarais and Cranmer, 2012). The PNF approach enables comparing groups of brain networks by assessing the topological consistency of key node sets within and between groups. However, it is a strictly inferential (and not modeling) approach, and thus precludes quantifying and predicting relationships between disease outcomes and network structure, and simulating network structure. Unlike the PNF, the MDMR framework allows controlling for confounding covariates in group comparisons via a "psuedo-F" statistic; however, it too lacks the ability to simulate networks or make predictions. It also fails to account for the dependence in connectivity patterns across voxels.

To address the limitations of the current methods, we propose a two-part mixed-effects modeling framework that allows modeling both the probability of a connection (presence/absence of an edge) and the strength of a connection if it exists. Models within this framework enable quantifying the relationship between an outcome (e.g., disease status) and connectivity patterns in the brain while reducing spurious correlations through inclusion of confounding covariates. The models provide a means to test for overall group differences in the strength and probability of network connections, group differences in network topology, and individual edge differences (edge covariates can be easily implemented in the model) while accounting for the complex dependence structures of the networks. They also enable prediction about an outcome based on connectivity structure and vice versa, simulating networks to gain a better understanding of normal ranges of topological variability, and thresholding networks leveraging group information. In short, this multivariate statistical and network scientific fusion approach allows going beyond just reporting an omnibus group comparison p-value and enables a more thorough examination of system level properties.

For the following discussion of the two-part mixed-effects modeling framework, we describe the motivating data concerning age-related cognitive decline in Section 2. We detail our modeling approach and its utility in Section 3 and use the aging data to illustrate the use of the proposed framework in Section 4. We conclude with a summary discussion including planned future research in Section 5.

## 2. MOTIVATING EXAMPLE

Our data come from a prior study that aimed to assess the neurological underpinnings of age-related cognitive decline by examining the effects of aging on the integration of sensory information (Hugenschmidt et al. 2009). The study has two age groups, healthy young adults aged $27 \pm 5.8$ years old (n=20) and healthy older adults aged $73 \pm 6.6$ years old (n=19). Three separate conditions of fMRI scans were used, resting, visual (viewing of a silent movie), and multisensory (MS) (visual and auditory - movie with sound), each lasting 5.6 minutes. Further details about these conditions along with additional network analyses can be found in a previous publication (Moussa et al., 2011). For each fMRI scan, blood-oxygen-level dependence (BOLD) contrast was measured using a 1.5T MRI scanner and a whole-brain gradient echo- planar imaging (EPI) sequence with the following parameters: 200 volumes with 24 contiguous slices per volume; slice thickness = 5.0 mm; in-plane resolution of 3.75 mm $\times$ 3.75 mm; TR = 1700ms.

To process the data, functional scans were normalized to standard brain space with a 4 x 4 x 5 mm voxel size. Data were band pass filtered (0.00765-0.068 Hz), and motion parameters, global signal, and mean white matter (WM) and cerebral spinal fluid (CSF) signals were regressed from the imaging data. Brain networks for each participant were then constructed by calculating Pearson correlation coefficients between the time courses of all node pairs adjusted for motion and physiological noises (see Hayasaka and Laurienti, 2010 for further details). These node time courses were obtained by averaging the voxel time courses in the 90 distinct anatomical regions (90 ROIs-Regions of Interest) defined by the Automated Anatomical Labeling atlas (AAL; Tzourio-Mazoyer et al., 2002). Refer to the Introduction and Figure 1 for further discussion on brain network generation.

## 3. TWO-PART MIXED MODELING FRAMEWORK FOR WEIGHTED BRAIN NETWORKS

### 3.1 Definition

Given that we have positively weighted networks, with negative weights set to 0/no connection (for reasons noted in Telesford et al., 2011; Cao et al., 2014; and others), a two-part mixed-effects model will be employed in order to be able to model both the probability of a connection (presence/absence) and the strength of a connection if it exists. Several two-part models have been proposed in the literature for a variety of applications (Albert and Shen, 2005; Tooze et al., 2002). However, they have yet to be developed for networks in general or, more specifically, for brain networks.

Let $Y_{ijk}$ represent the *strength* of the connection (quantified as the correlation in our case) and $R_{ijk}$ indicate whether a connection is present (*presence* variable) between node $j$ and node $k$ for the $i^{\text{th}}$ subject. Thus, $R_{ijk} = 0$ if $Y_{ijk} = 0$, and $R_{ijk} = 1$ if $Y_{ijk} > 0$ with conditional probabilities

$$P(R_{ijk} = r_{ijk}|\boldsymbol{\beta}_r; \ \boldsymbol{b}_{ri}) = \begin{cases} 1 - p_{ijk}(\boldsymbol{\beta}_r; \ \boldsymbol{b}_{ri}) & \text{if } r_{ijk} = 0 \\ p_{ijk}(\boldsymbol{\beta}_r; \ \boldsymbol{b}_{ri}) & \text{if } r_{ijk} = 1, \end{cases} \tag{1}$$

where $\boldsymbol{\beta}_r$ is a vector of population parameters (fixed effects) that relate the probability of a connection to a set of covariates ($\boldsymbol{X}_{ijk}$) for each subject and nodal pair (dyad), and $\boldsymbol{b}_{ri}$ is a vector of subject- and node-specific parameters (random effects) that capture how this relationship varies about the population average ($\boldsymbol{\beta}_r$) by subject and node ($\boldsymbol{Z}_{ijk}$). Hence, $p_{ijk}(\boldsymbol{\beta}_r; \ \boldsymbol{b}_{ri})$ is the probability of a connection between nodes $j$ and $k$ for subject $i$. We then have the following logistic mixed model (part I model) for the probability of this connection:

$$logit(p_{ijk}(\boldsymbol{\beta}_r; \ \boldsymbol{b}_{ri})) = \boldsymbol{X}'_{ijk}\boldsymbol{\beta}_r + \boldsymbol{Z}'_{ijk}\boldsymbol{b}_{ri}. \tag{2}$$

For the part II model, which aims to model the strength of a connection given that there is one, we let $S_{ijk} = [Y_{ijk}|R_{ijk} = 1]$. In our case, the $S_{ijk}$ will be the values of the correlation coefficients between nodes $j$ and $k$ for subject $i$. We can then use Fisher's Z-transform, denoted as $FZT$, and assume normality (we have empirically observed normality in strength distributions) for the following mixed model (part II model)

$$FZT(S_{ijk}(\boldsymbol{\beta}_s; \ \boldsymbol{b}_{si})) = \boldsymbol{X}'_{ijk}\boldsymbol{\beta}_s + \boldsymbol{Z}'_{ijk}\boldsymbol{b}_{si} + e_{ijk}, \tag{3}$$

where $\boldsymbol{\beta}_s$ is a vector of population parameters that relate the strength of a connection to the same set of covariates ($\boldsymbol{X}_{ijk}$) for each subject and nodal pair (dyad), $\boldsymbol{b}_{si}$ is a vector of subject- and node-specific parameters that capture how this relationship varies about the population average ($\boldsymbol{\beta}_s$) by subject and node ($\boldsymbol{Z}_{ijk}$), and $e_{ijk}$ accounts for the random noise in the connection strength of nodes $j$ and $k$ for subject $i$.

The covariates ($\boldsymbol{X}_{ijk}$) used to explain and predict both the presence and strength of connection are 1) $Net$: the average of the following network measures (categorized in Table 1 and further detailed in Simpson et al., 2013a and Rubinov and Sporns, 2010) in each dyad: Clustering ($C$), Global Efficiency ($Eglob$), Degree ($k$) (difference instead of average to capture "assortativity"), Modularity ($Q$), and Leverage Centrality ($l$); 2) $COI$:

Covariate of Interest (Age Group in our case); 3) $Int$: Interactions of the Covariate of Interest with the metrics in 1) and Sex; and 4) $Con$: Confounders (Sex, Years of Education, Spatial Distance (between nodes), and the square of Spatial Distance in our case). Thus, we can decompose $\boldsymbol{\beta}_r$ and $\boldsymbol{\beta}_s$ into $\boldsymbol{\beta}_r = [\,\beta_{r,0}\ \boldsymbol{\beta}_{r,net}\ \beta_{r,coi}\ \boldsymbol{\beta}_{r,int}\ \boldsymbol{\beta}_{r,con}]'$ and $\boldsymbol{\beta}_s = [\,\beta_{s,0}\ \boldsymbol{\beta}_{s,net}\ \beta_{s,coi}\ \boldsymbol{\beta}_{s,int}\ \boldsymbol{\beta}_{s,con}]'$ to correspond with the population intercepts and these covariates. For the random-effects vectors we have that $\boldsymbol{b}_{ri} = [\,b_{ri,0}\ \boldsymbol{b}_{ri,net}\ \boldsymbol{b}_{ri,dist}\ \boldsymbol{\delta}_{ri,j}\ \boldsymbol{\delta}_{ri,k}]'$ and $\boldsymbol{b}_{si} = [\,b_{si,0}\ \boldsymbol{b}_{si,net}\ \boldsymbol{b}_{si,dist}\ \boldsymbol{\delta}_{si,j}\ \boldsymbol{\delta}_{si,k}]'$, where $b_{ri,0}$ and $b_{si,0}$ quantify the deviation of subject-specific intercepts from the population intercepts ($\beta_{r,0}$ and $\beta_{s,0}$), $\boldsymbol{b}_{ri,net}$ and $\boldsymbol{b}_{si,net}$ contain the subject-specific parameters that capture how much the relationships between the network measures in 1) and the *presence* and *strength* of a connection vary about the population relationships ($\boldsymbol{\beta}_{r,net}$ and $\boldsymbol{\beta}_{s,net}$) respectively, $\boldsymbol{b}_{ri,dist}$ and $\boldsymbol{b}_{si,dist}$ contain the subject-specific parameters that capture how much the relationship between spatial distance and the *presence* and *strength* of a connection vary about the population relationships respectively, $\boldsymbol{\delta}_{ri,j}$ and $\boldsymbol{\delta}_{si,j}$ contain nodal-specific parameters that represent the propensity for node $j$ (of the given dyad) to be connected and the magnitude of its connections respectively, and $\boldsymbol{\delta}_{ri,k}$ and $\boldsymbol{\delta}_{si,k}$ contain nodal-specific parameters that represent the propensity for node $k$ (of the given dyad) to be connected and the magnitude of its connections respectively. In general, additional covariates can also be incorporated as guided by the biological context. We assume that $\boldsymbol{b}_{ri}$, $\boldsymbol{b}_{si}$, and $\boldsymbol{e}_i = \{e_{ijk}\}$ are normally distributed and mutually independent, with variance component covariance structures for $\boldsymbol{b}_{ri}$ and $\boldsymbol{b}_{si}$, and the standard conditional independence structure for $\boldsymbol{e}_i$. That is, $\boldsymbol{b}_{ri} \sim N(\boldsymbol{0}, \boldsymbol{\Sigma}_{ri}(\boldsymbol{\tau}_r) = \text{diag}(\boldsymbol{\tau}_r))$ where $\boldsymbol{\tau}_r = (\sigma_{r,0}^2,\ \boldsymbol{\sigma}_{r,net}^2,\ \boldsymbol{\sigma}_{r,dist}^2,\ \sigma_{r,node1}^2,\ \sigma_{r,node2}^2 \dots,\ \sigma_{r,node90}^2)'$ and $\boldsymbol{b}_{si} \sim N(\boldsymbol{0}, \boldsymbol{\Sigma}_{si}(\boldsymbol{\tau}_s) = \text{diag}(\boldsymbol{\tau}_s))$ where $\boldsymbol{\tau}_r = (\sigma_{s,0}^2,\ \boldsymbol{\sigma}_{s,net}^2,\ \boldsymbol{\sigma}_{s,dist}^2,\ \sigma_{s,node1}^2,\ \sigma_{s,node2}^2 \dots,\ \sigma_{s,node90}^2)'$ are the $q(=98) \times 1$ vectors of variances for each element of the random effects vectors, and $\boldsymbol{e}_i \sim N(\boldsymbol{0}, \boldsymbol{\Sigma}_{ei} = \sigma^2 \boldsymbol{I})$. Parameter estimation is conducted via restricted pseudo-likelihood (Wolfinger and O'Connell, 1993) with the residual approximation of the $F$-test for a Wald statistic employed for inference. However, the model can also be imbedded within other appropriate estimation and inferential approaches.

As alluded to in the Introduction, and detailed in the next section, this modeling framework allows *explaining* the relationship between covariates and network connectivity, *comparing* network connectivity among groups, *predicting* network connectivity based on participant and nodal characteristics, *simulating* networks as a means of assessing goodness-of-fit (GOF) and gaining a better understanding of network topological variability, and *thresholding* networks leveraging group information. For explaining, comparing, and thresholding, we remove random effects with estimated variance components equal to zero from the model detailed above in order to preserve power and control test size (Littell, 2006). For predicting and simulating networks we employ the full model as our GOF assessments have shown that it better captures network topology.

## 3.2 Framework Utility

### 3.2.1 Explain

Our framework allows explaining (quantifying) the relationship between endogenous network features and the probability and strength of a connection between nodes (brain areas) via estimation of $\beta_{r,net}$ and $\beta_{s,net}$, the relationship between exogenous covariate(s) of interest and confounders and the probability and strength of connections via estimation of $\beta_{r,coi}$, $\beta_{r,con}$, $\beta_{s,coi}$, and $\beta_{s,con}$, and if (and how) the relationships between network features and the probability and strength of connections varies for different values of the covariate(s) of interest (between young and older adults in our case) via estimation of $\beta_{r,int}$ and $\beta_{s,int}$. Specific interpretations of fixed effect parameters (after centering continuous covariates) for our data context are given below.

$\beta_{r,0}$ and $\beta_{s,0}$ — The log odds of an edge existing and the average strength of that connection for dyads with average values for the network metrics and spatial distance from the network of a young male with the average educational attainment.

$\boldsymbol{\beta}_{r,net}$ and $\boldsymbol{\beta}_{s,net}$ — The change in the log odds of an edge existing and the average strength of that connection for a dyad with each unit increase in the given network metric from the networks of young adults.

$\beta_{r,age}$ and $\beta_{s,age}$ — The change in the log odds of an edge existing and the average strength of that connection for dyads from the networks of older males with average values for the network metrics.

$\beta_{r,sex}$ and $\beta_{s,sex}$ — The change in the log odds of an edge existing and the average strength of that connection for dyads from the networks of younger females.

$\beta_{r,educ}$ and $\beta_{s,educ}$ — The change in the log odds of edges existing and the average strength of those connections with each year increase in educational attainment.

$\beta_{r,dist}/\beta_{r,dist^2}$ and $\beta_{s,dist}/\beta_{s,dist^2}$ — The quadratic change in the log odds of an edge existing and the average strength of that connection with each mm (scaled to dm for model fit) increase in spatial distance between the two nodes of a given dyad.

$\boldsymbol{\beta}_{r,age\times net}$ and $\boldsymbol{\beta}_{s,age\times net}$ — The additional change (relative to $\boldsymbol{\beta}_{r,net}$ and $\boldsymbol{\beta}_{s,net}$) in the log odds of an edge existing and the average strength of that connection for a dyad with each unit increase in the given network metric from the networks of older adults.

$\beta_{r,age\times sex}$ and $\beta_{s,age\times sex}$ — The additional change (relative to $\beta_{r,sex}$ and $\beta_{s,sex}$) in the log odds of an edge existing and the average strength of that connection for dyads from the networks of older females.

*3.2.2 Compare*

In addition to simply quantifying the relationship between covariates and the presence and strength of connections while reducing spurious correlations through inclusion of confounding covariates, this framework also allows making formal statistical comparisons about connectivity, network structure, and edge properties between groups (e.g., young vs. old, healthy vs. diseased, etc.). For instance, below are interpretations of possible combinations of significant and non-significant parameters from the previous section for our comparison of young and older adults (NS-not significant, S-significant).

$\beta_{r/s,age}$(NS), $\beta_{r/s,age \times sex}$ (NS), $\boldsymbol{\beta}_{r/s,age \times net}$(NS): No overall connectivity differences (proportion and strength of edges) or topological differences (relationship between dyad characteristics and probability and strength of edges; i.e., connectivity differences that vary as a function of the value of the network metrics) between young and older adults. May still have differences in specific edges that can be assessed by including edge indicator variables in the model as desired.

$\beta_{r/s,age}$(NS), $\beta_{r/s,age \times sex}$ (NS), $\boldsymbol{\beta}_{r/s,age \times net}$(S): Connectivity differences between young and older adults vary (potentially compensatorily) by the values of the network metrics (are functions of the network metrics).

$\beta_{r/s,age}$(S), $\beta_{r/s,age \times sex}$ (NS), $\boldsymbol{\beta}_{r/s,age \times net}$(NS): Overall connectivity differences, but no topological differences between young and older adults.

$\beta_{r/s,age}$(S), $\beta_{r/s,age \times sex}$ (NS), $\boldsymbol{\beta}_{r/s,age \times net}$(S): Connectivity differences between young and older adults vary (potentially compensatorily) by the values of the network metrics (are functions of the network metrics).

### 3.2.3 Predict

Our approach enables making predictions about brain function in several ways:

1) Predictions about the *presence* and *strength* of connections between brain regions based on a disease or behavioral outcome of a participant can be made by "learning" the model parameters (estimating the $\beta$s) and then using this estimated model for predictions based on new data, thus providing a quantitative means to improve diagnoses, prognoses, and treatment planning;

2) Predictions about the topological relationships between network metrics and the *presence* and *strength* of connections based on participant characteristics;

3) Predictions about network structure and its variability via simulations of networks based on model fits (see Section *3.2.4*).

Given our research interest for the aging data, we focus on the latter two. In Section 4.3 we focus on 2) for the example data, with a discussion of making predictions about network structure and its variability reserved for sections 3.2.4 and 4.4.

### 3.2.4 Simulate

The two-part mixed modeling framework can also be used to simulate random realizations of networks that retain constitutive characteristics of the original networks. The fitted model yields both a probability mass function for the likelihood of network connections (Equation 2) and probability density function for the strength distribution of network connections (Equation 3) given the covariates. Consequently, group- and individual-level networks can be simulated from this joint distribution, by following the approach of Song et al. (2013) for example, providing a number of benefits delineated below.

1) Model goodness-of-fit (GOF) assessment(s) aligned with the research question(s) of interest: Simulations allow assessing many facets of model fit by comparing a set of descriptive metrics (means, medians, distributions, etc.) from the simulated networks with those of the observed networks. For good fitting models, the metrics of interest should closely match (Simpson et al., 2011).

2) Representative group-based network creation  with anatomical (dyadic) information incorporated: This simulation framework fits all networks simultaneously and thus it does not become more labor intensive to create representative networks with an increasing sample size as is the case with the ERGM based approach (Simpson et al., 2012). Also unlike the ERGM approach, it enables refining the process through the inclusion of additional biologically relevant variables and anatomical information.

3) Network variability assessment: Insight into biological variability can be gleaned via the distribution of possible brain networks produced from the simulations. These analyses can aid in elucidating network level features that play a role in various brain disorders. For instance, our aging study analyses showed that young subjects have more degree assortative networks during the visual and multisensory tasks. We can empirically examine how this difference affects the variability of simulated networks, thus helping to illuminate neurological mechanisms that lead to age-related cognitive decline (e.g., lack of interconnectivity among high degree brain areas leads to less stability in neuronal communication in the brains of older adults).

The validity of the results from 2) and 3) (as well as most other subsequent analyses) require a good fitting model, thus we focus on 1) here.

*3.2.5 Threshold*

As noted in Section 3.1, additional covariates can be included as guided by the questions of interest. As a consequence, this framework provides a powerful approach to network thresholding via inclusion of dyadic indicator variables. Given the sensitivity of network topology (and subsequent network analyses) to the thresholding process, approaches that refine this process are paramount (Simpson et al., 2013a; van Wijk et al., 2010). As opposed to most approaches which just use individual participant-level data to threshold a network (Telesford et al., 2011), our framework allows leveraging group-level data for this purpose. Better distinguishing between "true" weak connections and those resulting from noise may mitigate some of the challenges associated with weighted network analysis and provide more confidence in resulting conclusions (Simpson et al., 2013a).

Implementing the group-level thresholding approach requires adding indicator variables for the dyads of interest to the strength model defined in Equation 3 and then appropriately assessing the significance of the associated parameter. Adding interactions of the grouping variable (e.g., age group in our case) and the dyads of interest allows leveraging the data in a particular group, as opposed to all groups, in assessing the significance of a dyad. Non-significant dyads can be considered candidates for removal from the networks of participants with weak connections in those areas. Future work will examine this thresholding capability and assess its performance in the brain network context.

## 4. DATA APPLICATION AND RESULTS

We analyzed the aging study data discussed in Section 2 with the two-part mixed-effects modeling framework detailed in Section 3.1. All standard modeling assumptions and diagnostics were checked (Muller and Fetterman, 2002; Cheng et al., 2009). Here we illustrate the framework's ability to explain, compare, predict, and simulate. As noted earlier, future work will provide a thorough examination of its thresholding capabilities as this will require a paper in itself.

### 4.1 Explain

The resulting parameter estimates, along with the standard errors and p-values (based on the residual approximation of the $F$-test for a Wald statistic), associated with each of the fixed effect covariates for each scan condition are presented in Table 2. These estimates quantify (explain) the relationship between the endogenous network features and the probability and strength of a connection between nodes (brain areas), the relationship between age and the confounders (sex, years of education, spatial distance between regions, and the square of spatial distance) and the probability and strength of connections, and if (and how) the relationships between network features and the probability and strength of connections varies between young and older adults. See Section 3.2.1 for detailed interpretations of each parameter.

As evidenced by the magnitude of the estimates in the table relative to their standard errors, global efficiency (Functional Integration) and leverage centrality (Information Flow) play an important role in explaining the presence of connections between two brain regions across task conditions, while clustering (Functional Segregation) serves as a relatively important factor in the strength of these connections. Clustering also proves important in brain connectivity for young adults, but not older adults, during the visual and multisensory tasks. Spatial distance well predicts connection probability and strength across task conditions.

### 4.2 Compare

Below is a summary of the important inferential results from the analysis of the aging data gleaned from Table 2.

<u>Rest</u>: *Presence* − No Age Differences

*Strength* − **Old:** - Same as Young when $C$, $Eglob$, $l$, and $Q$ are equal to their averages
  - Increases and higher than Young when $C$, $Eglob$, and $l$ increase
  - Decreases and lower than Young when $Q$ increases

<u>Conclusion</u>: There are no age-related differences in the overall number of connections. The age-related differences in the overall strength of connections vary by dyadic clustering, global efficiency, and leverage centrality, and overall modularity. More specifically, older adults have stronger connections between nodes with higher clustering, global efficiency, and leverage centrality values, and weaker overall connection strength with higher overall modularity than young adults.

<u>Visual</u>: *Presence* − **Old:** - Same as Young when $C$, $Eglob$, $l$, and $k$ difference are equal to their averages
  - Increases and higher than Young when $k$ difference increases

- Decreases and lower than Young when $C$, $Eglob$, and $l$ increase

*Strength* $-$ **Old:** - Same as Young when $l$ is equal to its average
- Increases and higher than Young when $l$ increases

Conclusion: The age-related differences in the overall number of connections vary by dyadic clustering, global efficiency, leverage centrality, and degree difference. More specifically, older adults are less likely to have connections between nodes with higher clustering, global efficiency, and leverage centrality values, and more likely to have connections between nodes with a greater degree difference (less assortative) than younger adults. The age-related differences in the overall strength of connections vary by dyadic leverage centrality. More specifically, older adults have stronger connections between nodes with higher leverage centrality values than young adults.

Multi: *Presence* $-$ **Old:** - Same as Young when $C$ and $k$ difference are equal to their averages
- Increases and higher than Young when $k$ difference increases
- Decreases and lower than Young when $C$ increases

*Strength* $-$ **Old:** - Lower for Old Males when $Eglob$ and $l$ are equal to their averages
- Same as Young for Old Females when $Eglob$ and $l$ are equal to their averages
- Increases faster for Old than Young when $Eglob$ and $l$ increase

Conclusion: The age related differences in the overall number of connections vary by dyadic clustering and degree difference. More specifically, older adults are less likely to have connections between nodes with higher clustering, and more likely to have connections between nodes with a greater degree difference (less assortative) than young adults. The age-related differences in the overall strength of connections vary by dyadic global efficiency and leverage centrality. More specifically, connection strengths increase faster for older adults between nodes as dyadic global efficiency and leverage centrality values increase. Additionally, older women have more overall connection strength than older men.

These comparison results provide a fairly comprehensive appraisal of structural differences between young and older adults. While plausible biological interpretations would be quite speculative at this point, the less degree assortative network structure in older adults during the visual and multisensory tasks provides a more explicable finding. Degree assortativity implies the existence of a resilient core of interconnected high-degree hubs (highly connected brain regions). During the performance of a task, the high degree nodes are those that belong to task-relevant networks (Moussa et al., 2011; Rzucidlo et al., 2013; Ma et al., 2012; Cao et al., 2014). The loss of assortativity in the older adults during the tasks suggests that there is reduced connectivity within the task-relevant networks and greater connectivity to other network nodes. This finding is consistent with cognitive studies showing that older adults are more vulnerable to distraction when performing tasks (Alain and Woods, 1999; Darowski et al., 2008; Grady et al., 2006).

*4.3 Predict*

Here we illustrate the utility of the modeling framework for prediction with the aging study data by focusing on degree ($k$) difference. Predictions based on other network metrics can be done similarly. Figures 2-7 show the 95% prediction intervals for connection probability and strength as a function of degree difference in young and older participants at rest, during the visual task, and during the multisensory task. The 95% prediction interval provides a model-based definition of normal range for both groups. In contrast to confidence intervals which yield an (narrower) interval estimate for an average of participants, the prediction interval gives an interval estimate for a single participant. As can be seen from Figure 2, older adults have a higher predicted connection probability when two nodes have the same degree; however, this probability drops off at a slightly faster rate than for the young as the disparity between the degrees of two nodes increases at rest and during a visual task. That is, older adults are predicted to have brain networks that are more degree assortative than young adults at rest. Contrastingly, Figures 4 and 6 show that younger adults are predicted to have more degree assortative networks during visual and multisensory tasks given the faster decay of their connection probabilities. As noted in the previous section, this finding is consistent with the hypothesis that older adults tend to be more distractable, or be more cognitively vulnerable, while engaged in a task. As evidenced by Figures 3, 5, and 7, older adults have a lower predicted connection strength that parallels the decay of the young adults as the disparity between the degrees of two nodes increases at rest, during a visual task, and during a multisensory task. That is, an older adult is likely to have weaker coupling between brain region pairs than a young adult regardless of the regional connectivity patterns.

*4.4 Simulate*

To assess the goodness-of-fit of our model to the aging study data, and to illustrate the utility of our framework for simulating realistic brain networks, we simulated 100 networks based on the fitted models to the data at rest, during the visual task, and during the multisensory task. We then calculated several (weighted) descriptive metrics commonly used in the neuroimaging literature for the observed and simulated networks. Table 3 displays the results for clustering coefficient ($C$), global efficiency ($Eglob$), characteristic path length ($L$), mean nodal degree ($K$), leverage centrality ($l$), and modularity ($Q$). As evidenced by the results, the simulated networks are very similar to the observed networks in terms of clustering, path length, and global efficiency, and relatively similar for leverage centrality and degree across task conditions. While the simulated networks are not as similar for modularity, our method provides a huge step forward methodologically given that, to our knowledge, it is the only approach that allows the simulation of weighted networks that capture several important topological properties simultaneously. Moreover, the framework serves as a foundation upon which future work can further refine network simulation accuracy.

## 5. DISCUSSION

The recent explosion of brain network analyses has led to a paradigm shift in neuroscience; however, the statistical methods needed to draw deeper biological insights from these analyses have lagged behind. Our two-part mixed-effects modeling framework fills this void and provides a comprehensive approach to studying system level brain properties to further our understanding of normal and abnormal brain function as illustrated by our analysis of the aging study data. As detailed in Section 3.2, the approach allows *explaining* the relationship between covariates and network connectivity, *comparing* network connectivity among groups, *predicting* network connectivity based on participant and nodal characteristics, *simulating* networks as a means of assessing goodness-of-fit (GOF) and gaining a better understanding of network topological variability, and *thresholding* networks leveraging group information.

Future work within this framework will focus on further demonstrating its simulation capabilities and thoroughly assessing its utility as a network thresholding aid as discussed in Section 3.2.5. Generating accurate and informative representative group-based networks via simulation provides one avenue for inquiry that has been underexplored in the literature (Simpson et al., 2012). Additionally, creating signature networks via individual and group-level simulations and examining the subsequent distributions of these generated networks may prove useful in disease prediction. For example, we plan to apply our approach to data from the Alzheimer's Disease Neuroimaging Initiative (ADNI) to determine if simulated networks and their distributions at baseline are predictive of cognitive decline at follow up.

Modifications and extensions to our framework also provide opportunities for future research. Longitudinal (or multitask) brain network modeling remains an untapped area in need of development (Simpson et al., 2013a). Extending our static modeling approach to this domain via incorporation of time-dependent random effects or Kronecker product covariance modeling (Galecki, 1994; Naik and Rao, 2001; Simpson, 2010; Simpson et al., 2014a; Simpson et al., 2014b) provides a potential solution. For the latter, following the notation in equations 2 and 3, we could assume that the random error is distributed $e_i \sim N(\mathbf{0}, \mathbf{\Sigma}_{ei} = \sigma^2(task)[\mathbf{\Gamma} \otimes \mathbf{\Omega}])$, where $\sigma^2(task)$ is the variability of connections strengths across the network dyads and varies by task, $\mathbf{\Gamma}$ contains the correlations between tasks, and $\mathbf{\Omega}$ contains the correlations between connection strengths of dyads which may be modeled explicitly or assumed to have the standard conditional independence structure (i.e., $\mathbf{\Omega} = \sigma^2 \mathbf{I}$). Specification of the structure for $\mathbf{\Gamma}$ should be informed by the study design and examination of observed correlations (Simpson et al., 2013a). Incorporating negative connections (i.e., negatively correlated nodes/brain regions) into static or longitudinal brain network analyses will also prove important. The lack of metrics for quantifying functional segregation and integration (e.g., $C$ and $Eglob$) with negatively weighted edges remains the current limiting factor.

Our framework is situated at the interface of statistical, network, and brain science, providing a synergistic analytic foundation for whole-brain network data. As mentioned above, this flexible approach can be modified and extended to refine its utility. It provides a step toward filling the methodological needs of the emerging area of brain network analysis and will engender deeper insights into the complex neurobiological interactions and changes that occur in many brain diseases and disorders.

## ACKNOWLEDGEMENTS

This work was supported by NIBIB K25 EB012236-01A1 (Simpson), and Wake Forest Older Americans Independence Center (P30 21332) and the Sticht Center on Aging (Laurienti). The authors thank F. DuBois Bowman from the Department of Biostatistics at Columbia University for his insightful conversations and guidance.

Table 1. Explanatory network metrics by category

| Category | Metric(s) |
| --- | --- |
| 1) Functional Segregation | Clustering Coefficient |
| 2) Functional Integration | Global Efficiency |
| 3) Resilience | Degree Difference |
| 4) Centrality and Information Flow | Leverage Centrality |
| 5) Community Structure | Modularity |

Table 2. Aging Data: Estimates, Standard Errors (SE), and P-values

| Parameter | Rest | | | Visual | | | Multisensory | | |
|---|---|---|---|---|---|---|---|---|---|
| | Estimate | SE | P-value | Estimate | SE | P-value | Estimate | SE | P-value |
| $\beta_{r,0}$ | -0.3141 | 0.0569 | < 0.0001 | -0.2905 | 0.0534 | < 0.0001 | -0.0585 | 0.0634 | 0.3556 |
| $\beta_{r,C}$ | 0.7807 | 0.3424 | 0.0226 | 6.7083 | 0.8969 | < 0.0001 | 13.3638 | 1.6908 | < 0.0001 |
| $\beta_{r,Eglob}$ | 32.6231 | 2.3322 | < 0.0001 | 34.7897 | 2.3725 | < 0.0001 | 32.8996 | 1.9054 | < 0.0001 |
| $\beta_{r,k}$ | -1.4442 | 0.1522 | < 0.0001 | -1.7862 | 0.1410 | < 0.0001 | -2.0598 | 0.1361 | < 0.0001 |
| $\beta_{r,Q}$ | -0.7345 | 1.1361 | 0.5179 | -0.8004 | 1.2082 | 0.5077 | -3.1672 | 1.5257 | 0.0379 |
| $\beta_{r,l}$ | 1.1598 | 0.0785 | < 0.0001 | 1.4156 | 0.0792 | < 0.0001 | 1.5381 | 0.0919 | < 0.0001 |
| $\beta_{r,age}$ | -0.0438 | 0.07733 | 0.5709 | 0.0884 | 0.0733 | 0.2279 | -0.0200 | 0.0866 | 0.8171 |
| $\beta_{r,sex}$ | -0.0085 | 0.0825 | 0.9178 | 0.0413 | 0.0725 | 0.5690 | -0.0479 | 0.0984 | 0.6265 |
| $\beta_{r,educ}$ | 0.0027 | 0.0103 | 0.7954 | -0.0007 | 0.0097 | 0.9467 | 0.0159 | 0.0130 | 0.2212 |
| $\beta_{r,dist}$ | -1.4266 | 0.0572 | < 0.0001 | -1.4593 | 0.0694 | < 0.0001 | -1.4419 | 0.0689 | < 0.0001 |
| $\beta_{r,dist^2}$ | 2.6558 | 0.1417 | < 0.0001 | 2.6648 | 0.1364 | < 0.0001 | 3.0032 | 0.2079 | < 0.0001 |
| $\beta_{r,age \times C}$ | 1.1249 | 0.7986 | 0.1590 | -6.3557 | 0.9239 | < 0.0001 | -10.2034 | 2.3529 | < 0.0001 |
| $\beta_{r,age \times Eglob}$ | -1.7255 | 3.3478 | 0.6063 | -7.8023 | 3.3969 | 0.0216 | -2.3070 | 2.7258 | 0.3974 |
| $\beta_{r,age \times k}$ | 0.2455 | 0.2185 | 0.2612 | 0.5473 | 0.2013 | 0.0066 | 0.7766 | 0.1943 | < 0.0001 |
| $\beta_{r,age \times Q}$ | 1.5858 | 1.5753 | 0.3141 | 2.1234 | 1.7684 | 0.2298 | 3.1948 | 1.9126 | 0.0948 |
| $\beta_{r,age \times l}$ | 0.0638 | 0.1154 | 0.5803 | -0.3755 | 0.1086 | 0.0005 | -0.2247 | 0.1324 | 0.0896 |
| $\beta_{r,age \times sex}$ | 0.1914 | 0.1145 | 0.0946 | -0.0544 | 0.1093 | 0.6187 | 0.1490 | 0.1353 | 0.2706 |
| $\beta_{s,0}$ | 0.2290 | 0.0091 | < 0.0001 | 0.2173 | 0.0133 | < 0.0001 | 0.2380 | 0.0130 | < 0.0001 |
| $\beta_{s,C}$ | 2.2940 | 0.2428 | < 0.0001 | 2.4104 | 0.3409 | < 0.0001 | 2.1541 | 0.2161 | < 0.0001 |
| $\beta_{s,Eglob}$ | 0.9534 | 0.1823 | < 0.0001 | 1.3179 | 0.1841 | < 0.0001 | 0.9801 | 0.2003 | < 0.0001 |
| $\beta_{s,k}$ | -0.2524 | 0.0153 | < 0.0001 | -0.0016 | 0.0244 | 0.9468 | -0.2523 | 0.0177 | < 0.0001 |
| $\beta_{s,Q}$ | -0.0373 | 0.1723 | 0.8285 | -0.4926 | 0.3018 | 0.1027 | -0.6634 | 0.3120 | 0.0335 |
| $\beta_{s,l}$ | -0.0036 | 0.0109 | 0.7426 | 0.0032 | 0.0147 | 0.8274 | -0.0164 | 0.0126 | 0.1948 |
| $\beta_{s,age}$ | -0.0106 | 0.0124 | 0.3907 | -0.0012 | 0.0180 | 0.9448 | -0.0429 | 0.0178 | 0.0157 |
| $\beta_{s,sex}$ | -0.0046 | 0.0132 | 0.7263 | 0.0209 | 0.0181 | 0.2496 | -0.0341 | 0.0202 | 0.0903 |
| $\beta_{s,educ}$ | -0.0009 | 0.0017 | 0.5814 | -0.0014 | 0.0024 | 0.5703 | 0.0028 | 0.0027 | 0.2922 |
| $\beta_{s,dist}$ | -0.1718 | 0.0054 | < 0.0001 | -0.1769 | 0.0055 | < 0.0001 | -0.1723 | 0.0054 | < 0.0001 |
| $\beta_{s,dist^2}$ | 0.3910 | 0.0118 | < 0.0001 | 0.3838 | 0.0141 | < 0.0001 | 0.3902 | 0.0148 | < 0.0001 |
| $\beta_{s,age \times C}$ | 0.7214 | 0.3494 | 0.0389 | 0.4342 | 0.4808 | 0.3665 | 0.1812 | 0.3009 | 0.5470 |
| $\beta_{s,age \times Eglob}$ | 0.7042 | 0.2623 | 0.0073 | 0.1866 | 0.2640 | 0.4797 | 0.6309 | 0.2864 | 0.0276 |
| $\beta_{s,age \times k}$ | 0.0009 | 0.0220 | 0.9662 | -0.0016 | 0.0244 | 0.9468 | 0.0024 | 0.0253 | 0.9234 |
| $\beta_{s,age \times Q}$ | -0.6957 | 0.2464 | 0.0048 | 0.2925 | 0.3990 | 0.4635 | -0.1478 | 0.3896 | 0.7044 |
| $\beta_{s,age \times l}$ | 0.0556 | 0.0157 | 0.0004 | 0.0464 | 0.0209 | 0.0265 | 0.0349 | 0.0183 | 0.0560 |
| $\beta_{s,age \times sex}$ | 0.0001 | 0.0184 | 0.9952 | -0.0312 | 0.0269 | 0.2471 | 0.0513 | 0.0276 | 0.0632 |

Table 3. Weighted network metrics of observed and simulated networks from the aging study data at rest.

| Condition | Metric | Observed (N=39) Mean (SE) | Simulated (N=100) Mean (SE) |
|---|---|---|---|
| Rest | Clustering coefficient ($C$) | 0.149 (0.001) | 0.155 (0.000) |
| | Global Efficiency ($Eglob$) | 0.231 (0.001) | 0.214 (0.000) |
| | Characteristic path length ($L$) | 4.677 (0.109) | 4.720 (0.004) |
| | Mean Nodal Degree ($K$) | 10.649 (0.055) | 12.747 (0.016) |
| | Leverage Centrality ($l$) | 2.678 (0.015) | 1.945 (0.003) |
| | Modularity ($Q$) | 0.342 (0.001) | 0.136 (0.000) |
| Visual | Clustering coefficient ($C$) | 0.150 (0.001) | 0.150 (0.000) |
| | Global Efficiency ($Eglob$) | 0.232 (0.001) | 0.205 (0.000) |
| | Characteristic path length ($L$) | 4.553 (0.073) | 4.992 (0.008) |
| | Mean Nodal Degree ($K$) | 10.656 (0.056) | 12.379 (0.025) |
| | Leverage Centrality ($l$) | 2.671 (0.015) | 2.155 (0.010) |
| | Modularity ($Q$) | 0.348 (0.001) | 0.136 (0.000) |
| Multisensory | Clustering coefficient ($C$) | 0.140 (0.001) | 0.165 (0.000) |
| | Global Efficiency ($Eglob$) | 0.230 (0.001) | 0.218 (0.000) |
| | Characteristic path length ($L$) | 4.431 (0.011) | 4.700 (0.008) |
| | Mean Nodal Degree ($K$) | 10.547 (0.052) | 13.830 (0.027) |
| | Leverage Centrality ($l$) | 2.862 (0.014) | 2.054 (0.006) |
| | Modularity ($Q$) | 0.327 (0.001) | 0.123 (0.000) |

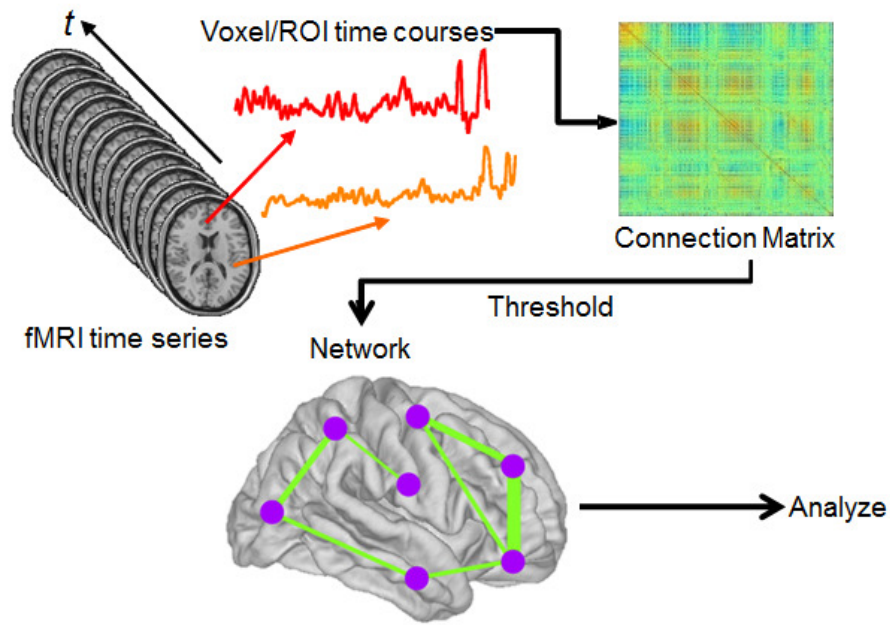

Figure 1. Schematic for generating brain networks from fMRI time series data (partially recreated from Simpson et al., 2013a and Fornito et al., 2012). Functional connectivity between brain areas is estimated based on time series pairs to produce a connection matrix. A threshold is commonly applied to the matrix to remove negative and/or "weak" connections.

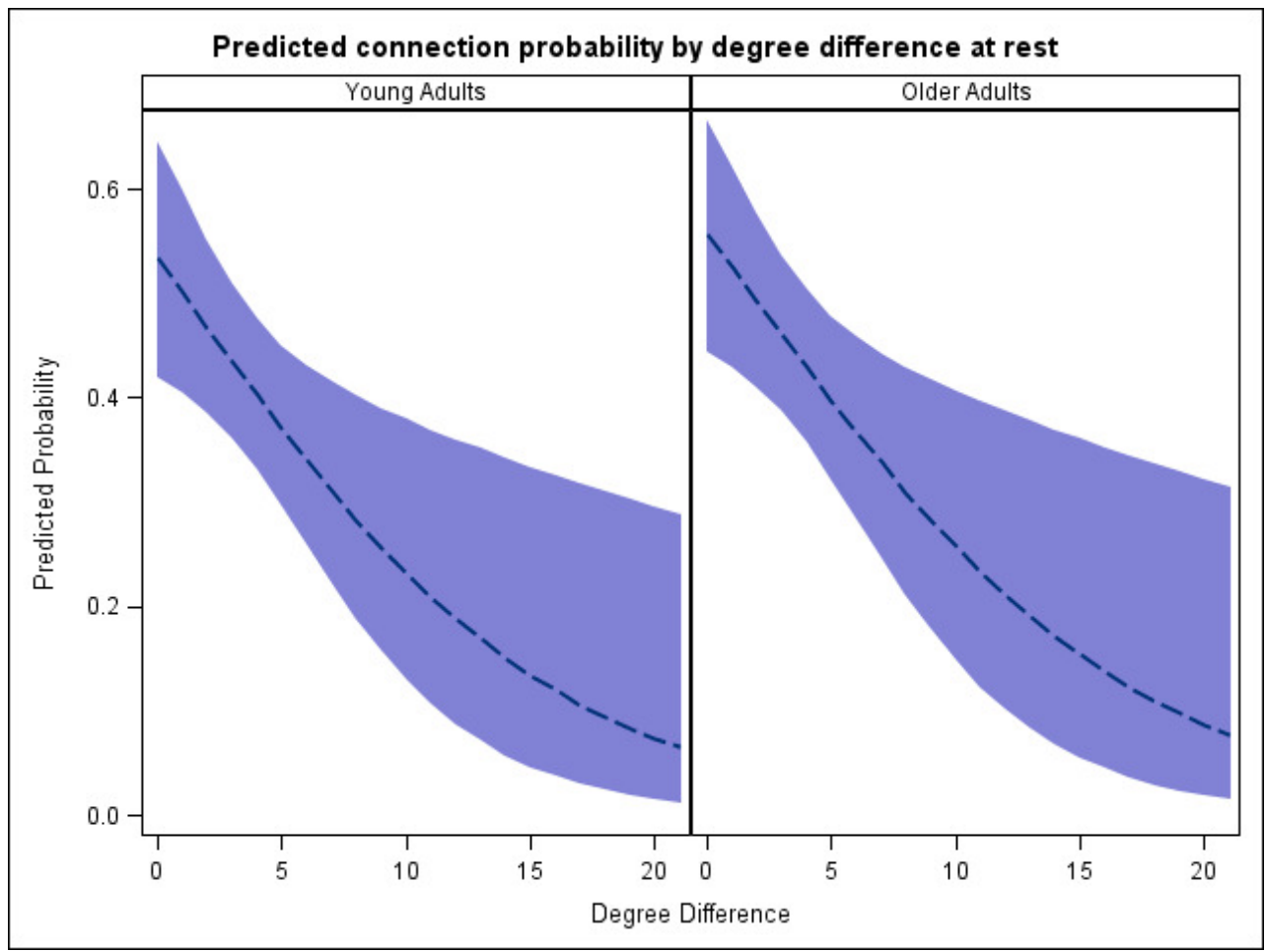

Figure 2. Prediction intervals for connection probability as a function of degree difference in young and older participants at rest.

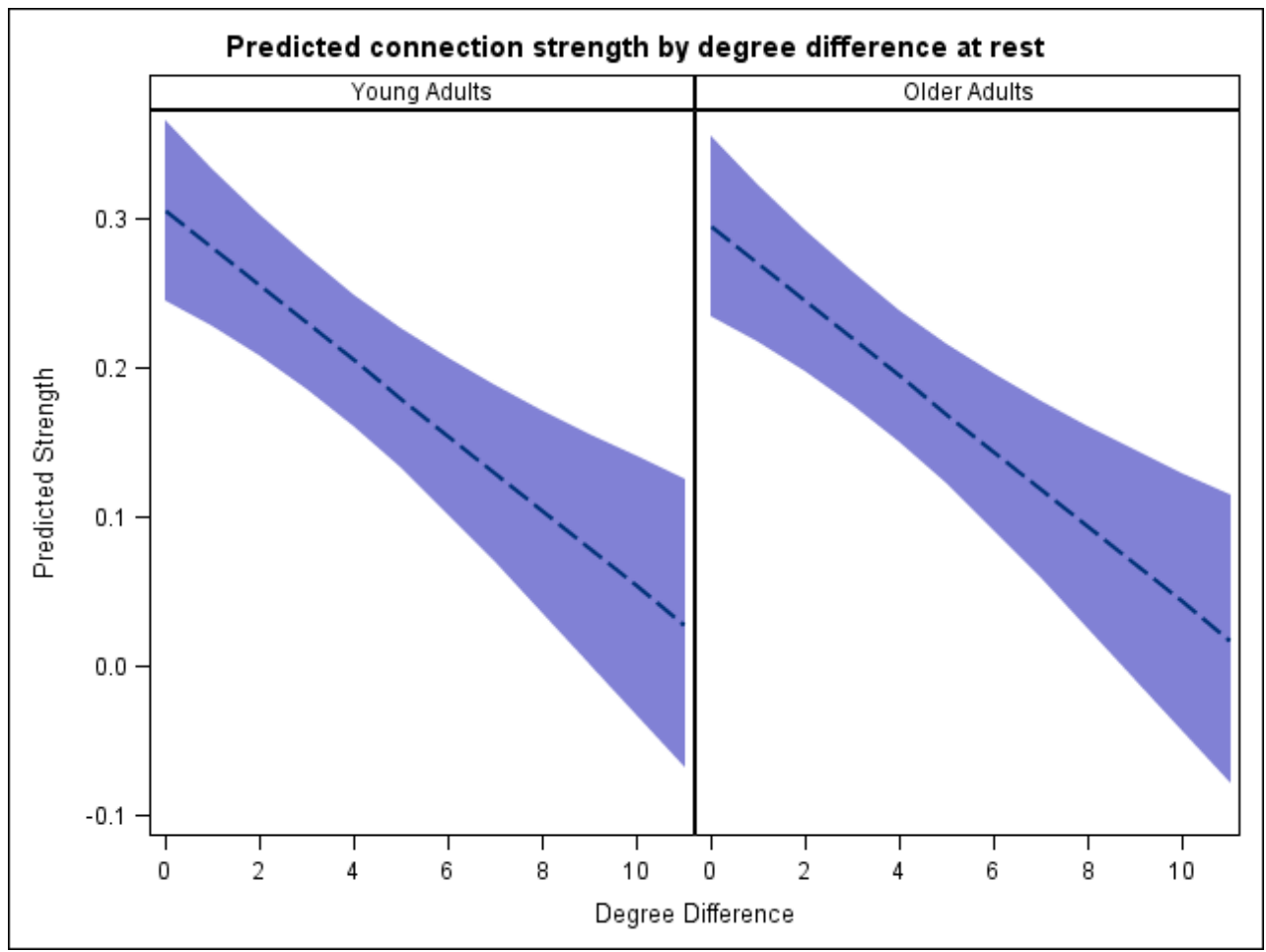

Figure 3. Prediction intervals for connection strength as a function of degree difference in young and older participants at rest.

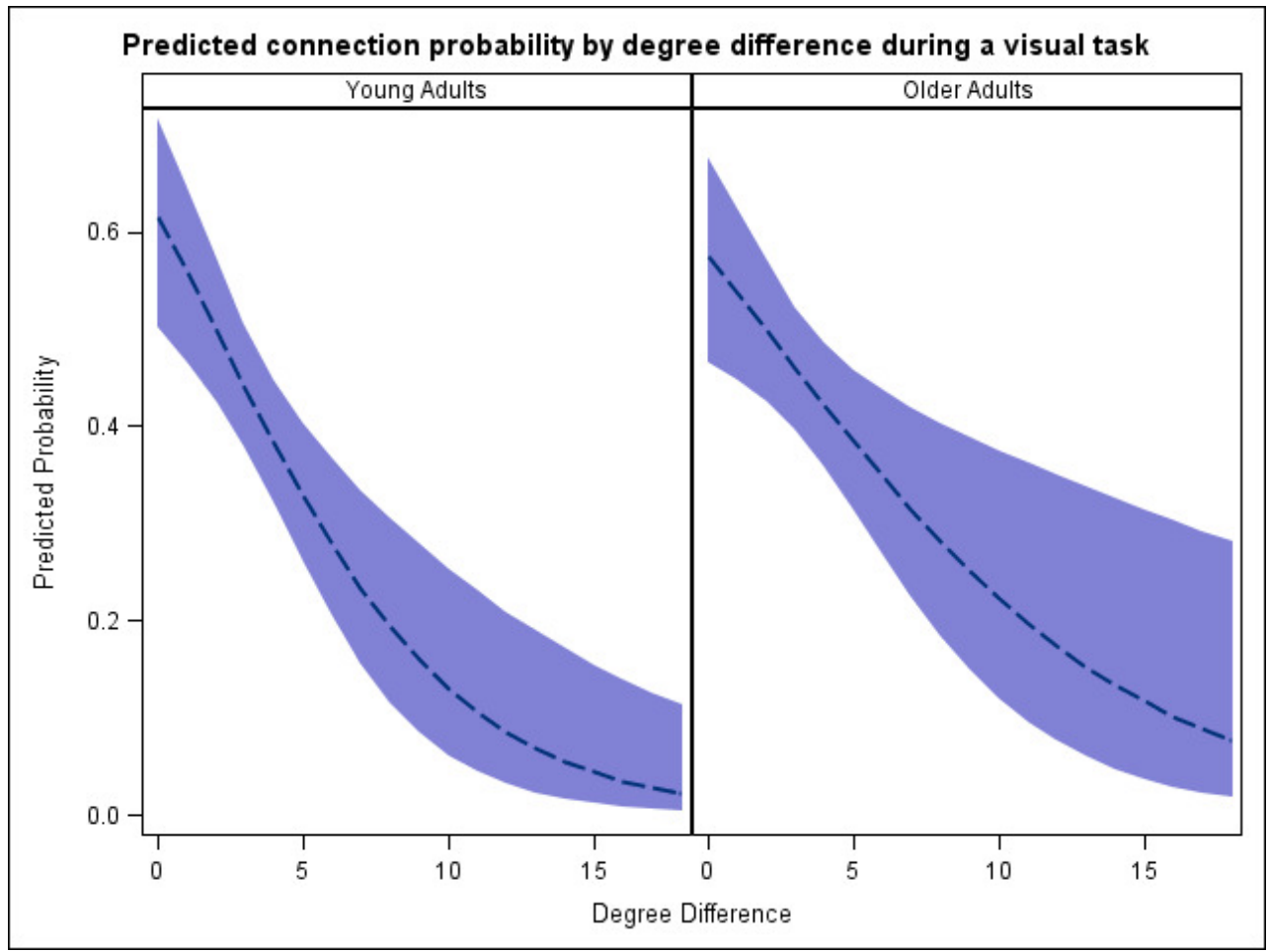

Figure 4. Prediction intervals for connection probability as a function of degree difference in young and older participants during a visual task.

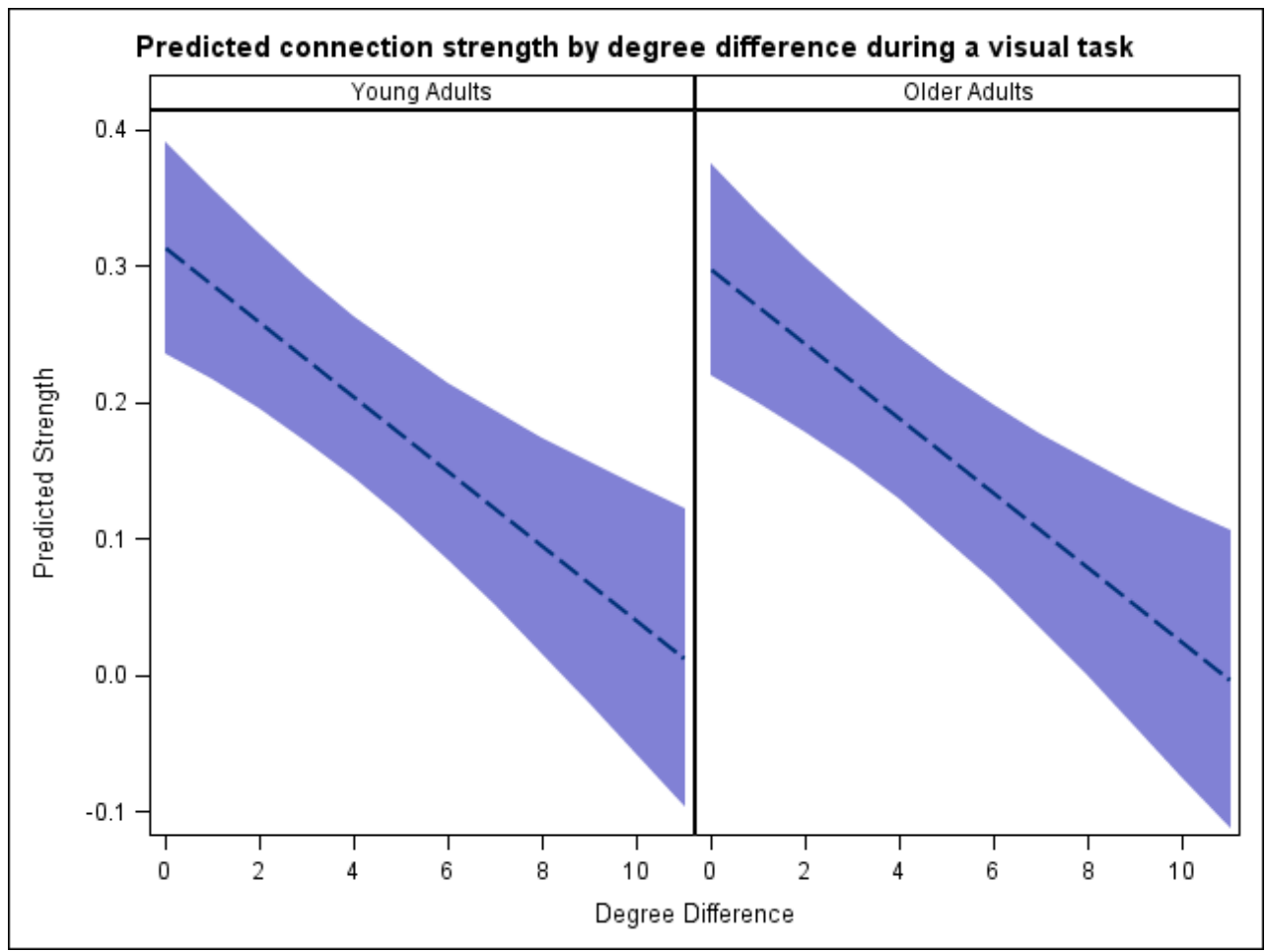

Figure 5. Prediction intervals for connection strength as a function of degree difference in young and older participants during a visual task.

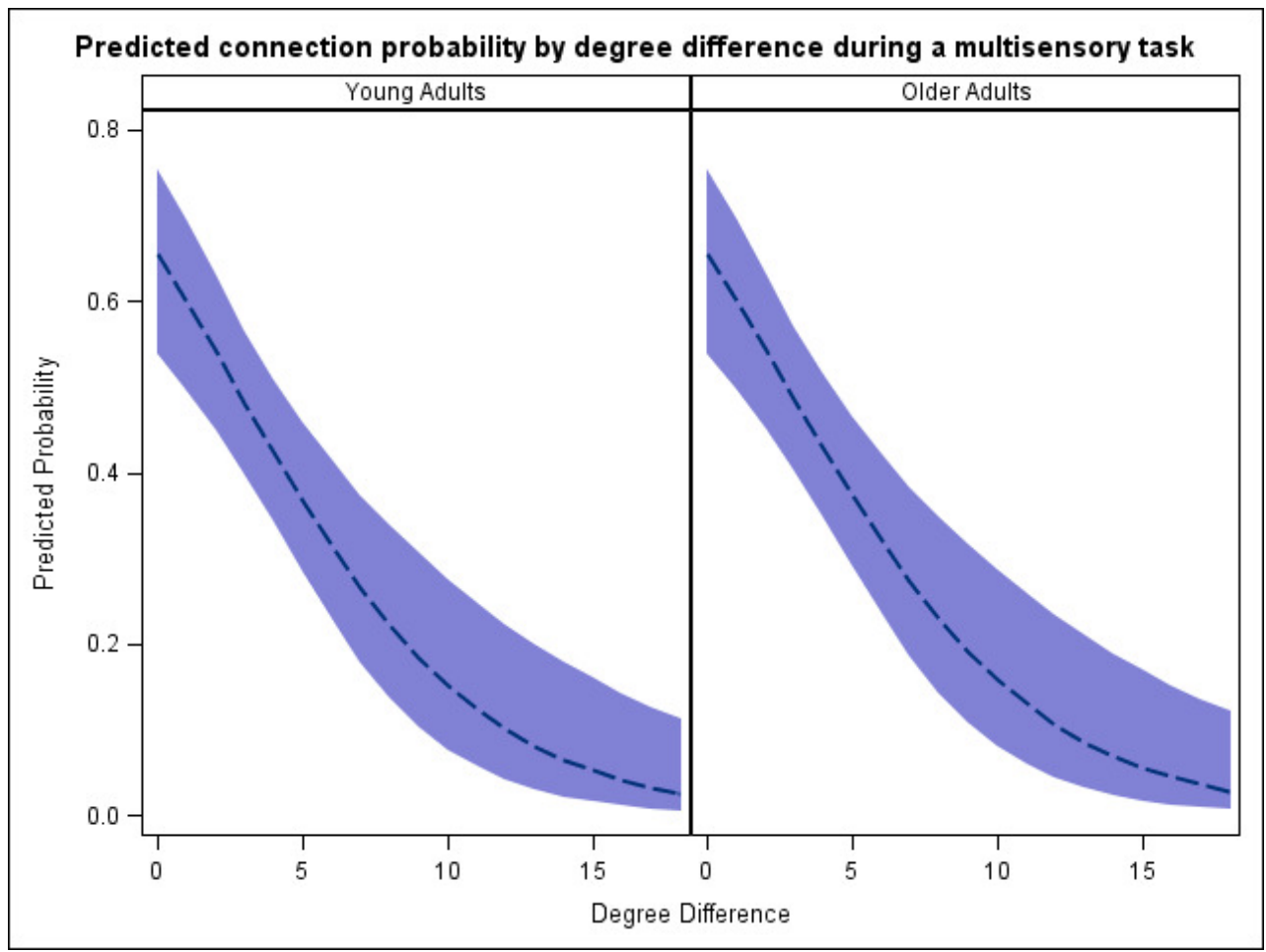

Figure 6. Prediction intervals for connection probability as a function of degree difference in young and older participants during a multisensory task.

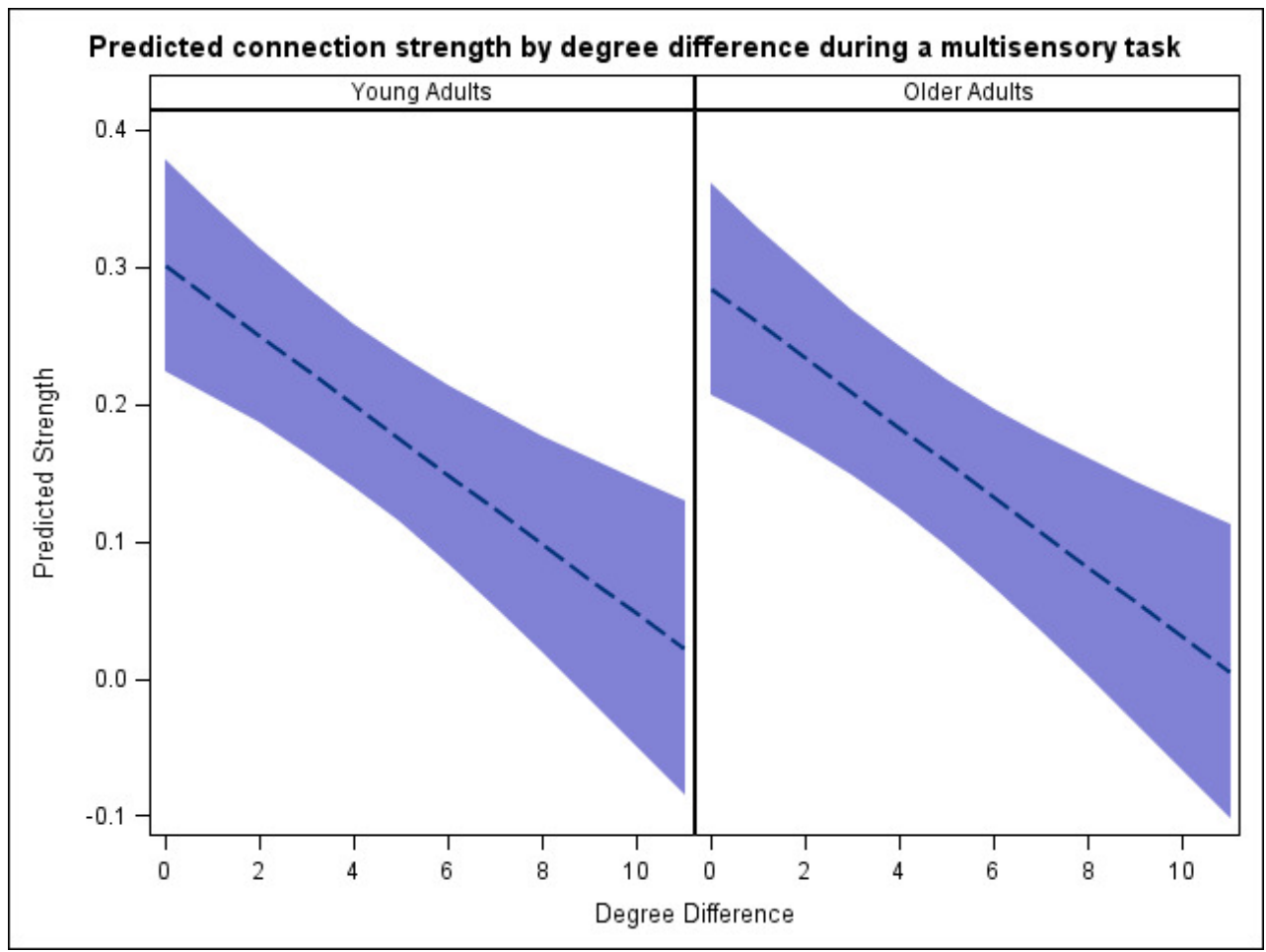

Figure 7. Prediction intervals for connection strength as a function of degree difference in young and older participants during a multisensory task.